# The Near-Infrared Fundamental Plane of Elliptical Galaxies

Michael A. Pahre[1,2], S. G. Djorgovski[1,3], and R. R. de Carvalho[1,4]

## ABSTRACT

We present results from a near-infrared $K$-band imaging survey of 59 elliptical galaxies in five nearby clusters. We measure photometric parameters for each galaxy using surface photometry and draw velocity dispersions from the literature. Three observables define a near-infrared Fundamental Plane (FP) of elliptical galaxies with $R_e \propto \sigma^{1.44\pm0.04} \Sigma_e^{-0.79\pm0.04}$. The scatter in the near-infrared relation is small at 16.5% in distance, which is equivalent to, or less than, the scatter of the optical FP. We suggest that the small deviation of the near-infrared FP relation from the optical FP is due to the reduction of metallicity effects in the near-infrared bandpass. While the small scatter of the optical FP could be consistent with compensating effects of age and metallicity, the similarly small scatter of the near-infrared FP is nearly independent of metallicity and hence places a strong constraint on possible age spreads among elliptical galaxies at every point along the FP. We suggest that the departure of the near-infrared FP from the pure virial form $R_e \propto \sigma^2 \Sigma_e^{-1}$, and the corresponding observed relation $(M/L) \propto M^{0.16\pm0.01}$, may be explained by slight systematic departures of the structure and dynamics of elliptical galaxies from a homology.

*Subject headings:* galaxies: clusters: general — galaxies: elliptical and lenticular, cD — galaxies: fundamental parameters — infrared: galaxies

[1]Palomar Observatory, California Institute of Technology, Pasadena, CA 91125; *Email:* map@astro.caltech.edu, george@deimos.caltech.edu, reinaldo@astro.caltech.edu

[2]Guest Observer, Las Campanas Observatory.

[3]Presidential Young Investigator.

[4]On Leave of absence from Observatório Nacional - CNPq - DAF.

## 1. Introduction

Elliptical galaxies form a two-parameter family, a two-dimensional Fundamental Plane (FP) within the multi-dimensional parameter space of several observables (Djorgovski & Davis 1987; Dressler et al. 1987; see also Kormendy & Djorgovski 1989, Djorgovski 1992, and references therein). Typically, one considers a parameter space of the half-light radius $r_e$, the mean surface brightness enclosed within that radius $\langle \mu \rangle_e$, and the central velocity dispersion $\sigma$. The FP correlations have a small scatter (of order 0.1 dex in $\log r_e$), which makes them potentially useful both as distance indicators and as a strong constraint on galaxy formation models. Explanation has been offered as to the origin of the FP, in particular its apparent deviation from the pure virial relation, primarily by invoking stellar populations variations among galaxies along the FP (Djorgovski 1988; Djorgovski & Santiago 1993, hereafter DS93; Renzini & Ciotti 1993). Substantial fine-tuning of variations in the IMF or mass-to-light ratio are required to remain consistent with the thinness at every location along the FP. Furthermore, the allowed range of mass-to-light ratio from one end of the FP to the other is small (Djorgovski & Davis 1987; Faber et al. 1987).

Variations in metallicity along the FP could give rise to the tilt of the FP in optical wavelengths, as line-blanketing effects in the metal-rich giant ellipticals would make their $\langle \mu \rangle_e$ fainter. Observations of the FP in the near-infrared could shed light on the origin of the FP by probing a wavelength region with a different sensitivity to stellar populations effects. Observations in the $K$-band window ($\lambda = 2.2 \mu$m) will trace old stellar populations and the bolometric luminosity, as this bandpass is far less metallicity-sensitive than optical wavelengths. Early results on the near-infrared FP (Recillas-Cruz et al. 1990, 1991; DS93) have all relied on photoelectric aperture photometry, whereas the optical FP comparison data utilized detailed surface photometry. The rapid development in near-infrared imaging detectors has only recently made a near-infrared imaging survey of this scale feasible.

The observations described in this *Letter* are preliminary results for part of an ongoing program to observe more than 250 early-type galaxies in eleven clusters. The complete details of the observational methods, reductions, calibrations, and comparisons with photometry from the literature will be given in a subsequent paper (Pahre et al. 1995b). These results follow preliminary results presented for the Virgo (Pahre et al. 1995a) and Perseus (Pahre et al. 1994) clusters individually, and for the entire five cluster sample (Djorgovski, Pahre, & de Carvalho 1995) which includes the Virgo, Perseus, Coma, Abell 194, and Abell 2634 clusters.

## 2. Observations

Nine galaxies in the Virgo cluster were observed in the $K_s$-band ($2.16 \mu$m) at the 2.5 m du Pont Telescope at the Las Campanas Observatory in 1993 March, and are described elsewhere (Pahre & Mould 1994). The observations of the other four clusters were taken in the $K_s$-band between 1994 October and 1995 March using a new near-infrared camera (Murphy et al. 1995) on the 1.52 m telescope at Palomar Observatory. This instrument is based on a NICMOS–3, $256 \times 256$ pixel array reimaged at 1:1 at the f/8.75 cassegrain focus, producing a $0.620''$ projected pixel size and a $159''$ field-of-view. The galaxy lists were mostly chosen from those of Lucey & Carter (1988) and Lucey et al. (1991a,b), except for the Perseus cluster, which was selected to include all 19 galaxies in Faber et al. (1989). Each galaxy was observed for $15 \times 60$ s on-source, separated into five different telescope pointings, and the sky was estimated from a similar total number of frames taken before and after each galaxy. All observations were taken during photometric conditions, with a mixture of the UKIRT faint standards (Casali & Hawarden 1992) and the new HST faint standards (S. E. Persson, private communication) being used for calibration. Usually between 5 and 15 standards were observed on each photometric night, which have a typical rms scatter less than 0.02 mag. The typical seeing was $\leq 1.5''$, with some observations having significantly better seeing than could be fully-sampled. We have compared aperture magnitudes measured for the Virgo and Coma galaxies with Persson, Frogel, & Aaronson (1979), and find mean offsets of $K_{our} - K_{PFA} = 0.00 \pm 0.01$ mag (Virgo) and $+0.02 \pm 0.02$ mag (Coma), with an rms scatter of 0.05 and 0.04 mag, respectively.

We have constructed surface brightness profiles for each galaxy by finding the best-fitting elliptical isophote $\mu$ in integer pixel steps in the semimajor axis $a$ using the ELLIPSE task in the STSDAS package of the IRAF image reduction software. Profiles at $a > 10$ pixels were rebinned to improve the SNR as described in Djorgovski (1985), and then corrections to the photometry for surface brightness (SB) dimming as $(1+z)^{-4}$, galactic extinction (taking $A_K = 0.085 A_B$ and the median $A_B$ estimate from Faber et al. 1989 for each cluster), and the k-correction of $+3.3z$ for the $K$-band (Persson et al. 1979). The mean SB internal to each isophote $\langle \mu \rangle$ was calculated by integrating the profile at each semimajor axis length. Half-light effective radii $r_e$ and the mean SB interior to that radius $\langle \mu \rangle_e$ were measured using fits of $\mu$ against $a^{1/4}$ under the assumption that all profiles followed a de Vaucouleur form. The apparent half-light radii $r_e$ were then transformed to metric size $R_e$ by assuming $H_0 = 75$ km s$^{-1}$. We have drawn velocity dispersion $\sigma$ data from the literature, giving equal weights to the work by Lucey & Carter (1988) or Lucey et al. (1991a,b) and a mean of all other measurements from the literature (as compiled by McElroy 1994, private communication). We note that using the Faber et al. (1989) velocity dispersions for the Coma cluster causes no significant change in the FP parameters described in §3. We have excluded all galaxies with $\sigma < 100$ km s$^{-1}$ or $r_e < 2''$, thereby avoiding issues of dwarf galaxies and seeing corrections. The galaxy sample is thus 59 galaxies out of the initial list of 83.

## 3. Constructing the Near-Infrared Fundamental Plane

Using the measurements of $r_e$, $\langle \mu \rangle_e$, and $\sigma$ described in §2, we have used bivariate least-squares analysis to find the optimal correlation between these FP variables. The methodology is as found in previous studies (Djorgovski & Davis 1987; de Carvalho & Djorgovski 1989; de Carvalho & Djorgovski 1992). In order to construct the FP itself, we first find the "mixing" value $b$ in the quantity $\log \sigma + b \langle \mu \rangle_e$ which has the highest correlation with $\log r_e$, and estimate the errors in $b$ from the "bootstrap" method. We then measure the slope $A$ and intercept $C$ for the FP relation

$$\log r_e (\text{arcsec}) = A \log \sigma - 0.4 B \langle \mu \rangle_e + C, \quad (1)$$

where $B = -2.5Ab$. The parameters $A$ and $C$ are measured using a linear least-squares relation by minimizing the variance orthogonal to the best-fitting line. We have measured the parameters $A$, $B$, and $C$ for all five clusters separately, and find that the values of $A$ and $B$ are consistent within the uncertainties. Taking the median values of $A$ and $B$, we then measured the median intercept $C'$ for each cluster, and used this value of $C'$ to transform each galaxy onto a metric scale (under the assumption that the Coma cluster has no peculiar velocity and $H_0 = 75$ km s$^{-1}$). We then measured the value for $A$ and $B$ for the entire sample simultaneously in metric units for $R_e$, and find that the entire dataset is best-fit by the $K$-band FP relation with $A = 1.435 \pm 0.040$, $B = -0.79 \pm 0.04$. This result is consistent with the median result from the individual clusters, and corresponds to the scaling relation $r_e \propto \sigma^{1.44 \pm 0.04} \Sigma_e^{-0.79 \pm 0.04}$ (where $\Sigma_e$ is the mean surface brightness within $r_e$ expressed in linear units, and $\langle \mu \rangle_e$ is the same quantity expressed in mag arcsec$^{-2}$). The results for the individual clusters are detailed in Table 1 and displayed in Figure 1.

We have calculated the quartile-estimated scatter in $\log R_e$ to be $Q\sigma = 0.072$ dex for the entire dataset, which corresponds to an error in distance of 16.5% per galaxy. This dispersion about the near-infrared FP is equivalent to, or smaller than, the 0.088 dex (Djorgovski & Davis 1987), 0.07 dex (de Carvalho & Djorgovski 1992, for the Faber et al. 1989 cluster sample), or 0.075 dex (Lucey et al. 1991b) found for the optical FP relation. Measurement errors are typically 0.04 dex in $\log r_e$, 0.017 mag in fitting $\langle \mu \rangle_e$, 0.02 dex due to sky-subtraction uncertainties, and 0.03 dex in $\log \sigma$, which suggests that much of the observed 0.072 dex thickness of the near-infrared FP can be accounted for by the 0.063 dex observational errors, resulting in an upper limit of 0.035 dex for the intrinsic thickness of the near-infrared FP.

Our results for the near-infrared FP are consistent with previous photoelectric photometry results. For the $K$-band, DS93 found $A = 1.5 \pm 0.1$ and $B = -0.78 \pm 0.17$ for the $r_e$ measurements of Faber et al. (1989) combined with the colors of Persson et al. (1979). Recillas-Cruz et al. (1990, 1991), found $A = 1.69 \pm 0.11$ and $B = -0.74 \pm 0.05$ for the Coma cluster and $A = 1.55 \pm 0.19$ and $B = -0.81 \pm 0.10$ for the Virgo cluster using their own $K$-band measurements.

We also investigated the Kormendy relation (see Kormendy & Djorgovski 1989 and references therein) of $\log R_e (\text{kpc}) = k \langle \mu \rangle_e + l$, which is an oblique projection of the FP. We assume the relative distances derived from the FP relation as described above in converting $r_e$ to $R_e$. We use a least-squares minimization of the orthogonal distance from a linear relation between these two variables. We fit each cluster independently and find that they are all well-fit by values for $k$ that are consistent within one-sigma of the mean (and median) value for $k$, therefore we can fit the entire sample to determine the mean relation. The best fit for the five clusters is $k = 0.332 \pm 0.023$ and $l = -5.090 \pm 0.022$, corresponding to the scaling relation $R_e \propto I_e^{-0.83 \pm 0.06}$. This relation has a quartile-estimated dispersion of $Q\sigma = 0.173$, corresponding to an uncertainty of 40% in distance per galaxy. This best-fitting Kormendy relation is also plotted in Figure 1. The near-infrared relation can be compared to the optical $r$-band relation of $R_e \propto I_e^{-0.85}$ (Kormendy & Djorgovski 1989). The near-infrared FP thus shows a 60% reduction in scatter over the near-infrared Kormendy relation.

## 4. Discussion and Conclusions

The best-fit to Eq. (1) of the near-infrared FP relation can be compared with its optical counterpart. The $V$-band relation is $A = 1.24 \pm 0.08$ and $B = -0.9 \pm 0.1$ (Lucey, Bower, & Ellis 1991, taking the mean of the Coma and Virgo cluster results) or $A = 1.25 \pm 0.07$ and $B = -0.80 \pm 0.03$ (the cluster sample of Faber et al. 1989, in the analysis of de Carvalho & Djorgovski 1992). The variation in the exponent of the FP (i.e. the $R_e \propto \sigma^A$ term) with bandpass was described by DS93, with the slope of the FP becoming steeper ($A$ increases) with increasing wavelength. We find that $A$ has changed by $+0.19 \pm 0.06$ dex from the $V$-band to the $K$-band, consistent with the DS93 result of $+0.29 \pm 0.11$.

The relationship in Eq. (1) departs, however, from that expected for a pure virial theorem, if light directly traces mass. The virial form of the FP is $R_e \propto \sigma^2 \Sigma_e^{-1}$, so the near-infrared FP formally differs from the virial relation at high confidence for the $A$ parameter. Faber et al. (1987) and Djorgovski (1988) argue that if elliptical galaxies form a homologous family, then the mass-to-light ratio enclosed by $R_e$ is a function of luminosity according to the scaling relation $(M/L)_e \propto M^\alpha$, where $\alpha = (2 - A)/(2 + A)$. [We note that the alternate form found in some studies is $(M/L)_e \propto L_e^\beta$, where $\beta = (2 - A)/2A$ and $\alpha = \beta/(1 + \beta)$.] The optical studies listed above result in $\alpha = 0.23 \pm 0.02$ in the $V$-band. The near-infrared result is $\alpha = 0.16 \pm 0.01$, which is inconsistent with zero to high significance. Since the near-infrared sample described in this *Letter* spans $\sim 4$ mag in luminosity, the least luminous galaxies have a mass-to-light ratio only 50% less than that of the largest galaxies. Our near-infrared result is also consistent with DS93 and Recillas-Cruz et al. (1990, 1991) who found $\alpha = 0.14 \pm 0.03$ and $0.13 \pm 0.05$, respectively. This dependence of mass-to-light ratio on mass is plotted in terms of observables in Figure 2, although we note (as does Djorgovski 1988) that this figure shows an oblique projection of the FP, prone to large cumulative and correlated errors. Nonetheless, the FP-derived value of $\alpha = 0.16$ is consistent with the data of Figure 2.

Explanations of the departure of the optical relation from zero often focused on line-blanketing effects (Faber et al. 1987) as there is a mass-metallicity relation along the elliptical galaxy sequence. But while the slope $A$ of the near-infrared FP could be explained by stellar populations, the small scatter of the relation would be more challenging. For example, by assuming a uniform mapping between the color-magnitude relation and the FP, we find that the $(V - K)$ color-magnitude relation for ellipticals (Bower, Lucey, & Ellis 1992) predicts a change in $(V - K)$ of $\sim 0.32$ mag for the 4 mag range in $K_T$ of our observations. In the Worthey (1994) stellar populations models for solar metallicity and a 12 Gyr age, the color-magnitude relation would require only a change of 0.02–0.04 dex in $(M/L)_{Bol}$ between the two extremes of the FP, which is less than the 0.3 dex required in the $K$-band, thus the tilt of the FP cannot be explained by metallicity effects alone. A variation in age between 8 Gyr and 17 Gyr along the sequence would produce a 0.25 dex variation in $(M/L)_{Bol}$, but this would require a synchronicity in all clusters in the formation of elliptical galaxies of a given mass, since the thickness of the FP at any point is only $< 5\%$ of its span in age. Another explanation for the variation in $(M/L)$ along the FP could be the velocity anisotropy, but DS93 have used correlations utilizing velocity anisotropy and showed that they follow similar scaling relations to those described above; they conclude that velocity anisotropy could not explain all of the $(M/L)$ dependence on mass. IMF variations may be partly responsible (DS93), but they require a fine-tuning which appears unlikely (Renzini & Ciotti 1993); the same may be said for a systematic variation in the dark matter content of elliptical galaxies.

Under the assumption that elliptical galaxies form a homology (in both their spatial and kinematic distributions), there is a unique mapping between the observed central velocity dispersion and the mean kinetic energy per unit mass, and between the de Vaucouleur radius and the mean harmonic radius (Djorgovski, de Carvalho, & Han 1988). It is well known, however, that more luminous ellipticals tend to be more anisotropic (Davies *et al.* 1983) and to have shallower surface brightness profiles (Schombert 1986; Caon, Capaccioli, & D'Onofrio 1993). In addition, in $N$-body simulations there appear to be velocity dispersion aperture effects which entail a nonhomologous velocity distribution at any fiducial radius for galaxies of the same mass (Capelato, de Carvalho, & Carlberg 1995). Simple analytical models of elliptical galaxies can be constructed (Hjorth & Madsen 1995), which would reproduce the necessary surface brightness profile departures from the homology (Burkert 1993). The observations in this *Letter* that $\alpha \neq 0$ in the $K$-band, suggests that the origin of much of the tilt of the FP (relative to constant $(M/L)$ and the pure virial theorem) can be explained by systematic deviations of the structure and dynamics of ellipticals from a homology. In other words, the mappings described above may be changing along the sequence of elliptical galaxies. It is thus possible that the slope of the near-infrared FP can be explained primarily by dynamical arguments. In such a scenario in which slight departures from a homology are present within the family of elliptical galaxies, the small thickness of the FP might place strong constraints on galaxy formation (Capelato *et al.* 1995). More clues about this could be afforded by simulations spanning a wider range of galaxy properties, or by observations of the FP at higher redshifts.

This work was supported in part by the NSF PYI award AST-9157412 to S. G. D., and Greenstein and Kingsley Fellowships to M. A. P. Many thanks are due to the builders of the new near-infrared camera, in particular David Murphy, Eric Persson, and Anand Sivaramakrishnan. The construction of the camera was supported in part by generous donations from the Perkin Fund, the Jeanne Rich Foundation, and the North American Rockwell Corporation. We would like to thank Jeremy Mould for use of the Virgo cluster data, and the director of Las Campanas

Observatory for allocating time for infrared imaging of elliptical galaxies. We also thank J. Kennefick, C. Fassnacht, and J. Larkin for doing some of the observing for this project.

Table 1. Individual Cluster Solutions for the Near-Infrared Fundamental Plane

| Cluster | $N$ | $A$ | $\pm$ | $B$ | $\pm$ | $C$ | $\pm$ | $Q\sigma$ | $C'$ | $Q\sigma$ |
|---|---|---|---|---|---|---|---|---|---|---|
| Abell 194  | 13 | 1.12 | 0.10 | -0.76 | 0.10 | -6.759 | 0.021 | 0.070 | -7.679 | 0.068 |
| Abell 2634 | 10 | 1.44 | 0.06 | -0.75 | 0.13 | -6.696 | 0.018 | 0.051 | -8.037 | 0.055 |
| Perseus    | 16 | 1.67 | 0.17 | -0.79 | 0.12 | -8.234 | 0.024 | 0.085 | -7.694 | 0.087 |
| Coma       | 12 | 1.29 | 0.08 | -0.71 | 0.06 | -6.993 | 0.016 | 0.053 | -7.814 | 0.065 |
| Virgo      |  8 | 1.62 | 0.06 | -0.93 | 0.05 | -8.436 | 0.012 | 0.036 | -7.083 | 0.055 |

---

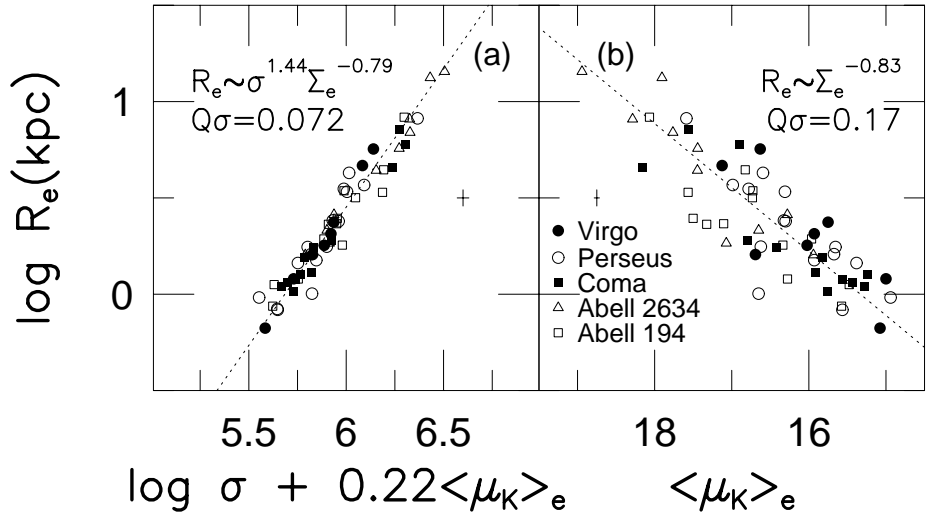

Fig. 1.— (a) The near-infrared FP of elliptical galaxies and (b) the Kormendy relation (an oblique projection of the FP) for the sample of 59 galaxies. The scatter is clearly reduced by the introduction of the central velocity dispersion term. The total scatter from the FP relation in units of $\log R_e$ is given as $Q\sigma = 0.072$, corresponding to a 16.5% uncertainty in distance per galaxy. Median error bars are also shown in each panel.

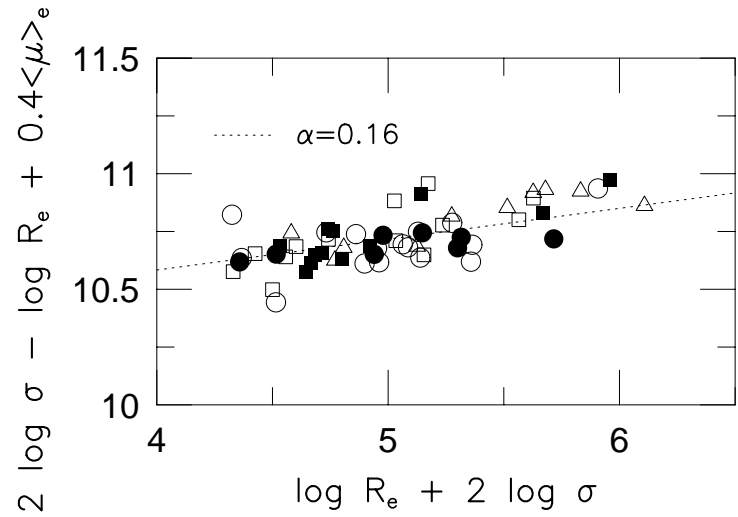

Fig. 2.— The variation of mass-to-light ratio (ordinate plus a constant) with mass (abscissa plus a constant) for the 59 galaxies in the sample. This diagram by itself is of limited utility in constraining $\alpha$, as the errors in both axes are both cumulative and correlated. The data are well-represented, however, by the FP-derived $\alpha = 0.16$, with a residual scatter of $Q\sigma = 0.11$ dex about this line.